\documentclass[reqno,11pt]{amsproc}  

\usepackage[utf8]{inputenc}
\usepackage{graphicx,float,epstopdf,cite,multirow}
\usepackage{amsmath,amsfonts,amssymb,amsaddr,enumitem}
\usepackage{adjustbox,tabularx,array,booktabs,ragged2e,siunitx}
\usepackage[a4paper, margin=0.8in,headsep=.4in]{geometry}

\usepackage{tikz}
\usetikzlibrary{arrows,positioning,patterns,shapes.geometric,calc}
\tikzset{>=stealth'}

\usepackage{pgfplots,pdfpages}
\usepackage{caption}
\usepackage{subcaption}

\definecolor{myblue}{rgb}{0.03,0.35,0.49}
\definecolor{dgreen}{rgb}{0.,0.6,0.}

\providecommand{\keywords}[1]
{
  \small	
  \textbf{\textit{Keywords---}} #1
}

\usepackage[round]{natbib}

\title[Effective poverty alleviation]{Effective alleviation of rural poverty depends on the interplay between productivity, nutrients, water and soil quality}

\author{Sonja Radosavljevic$^{1,*}$, L. Jamila Haider$^1$, Steven J. Lade$^1$, Maja Schlüter$^1$ }\thanks{$^*$ Corresponding author: Sonja Radosavljevic, sonja.radosavljevic@su.se  \\$^1$ Stockholm Resilience Centre, Stockholm University, Kräftriket 2B, 10691 Stockholm, Sweden\\  jamila.haider@su.se; steven.lade@su.se; maja.schlueter@su.se}

\begin{document}

\begin{abstract}
    Most of the world’s poorest people come from rural areas and depend on their local ecosystems for food production. Recent research has highlighted the importance of self-reinforcing dynamics between low soil quality and persistent poverty but little is known on how they affect poverty alleviation. We investigate how the intertwined dynamics of household assets, nutrients (especially phosphorus), water and soil quality influence food production and determine the conditions for escape from poverty for the rural poor. We have developed a suite of dynamic, multidimensional poverty trap models of households that combine economic aspects of growth with ecological dynamics of soil quality, water and nutrient flows to analyze the effectiveness of common poverty alleviation strategies such as intensification through agrochemical inputs, diversification of energy sources and conservation tillage. Our results show that (i) agrochemical inputs can reinforce poverty by degrading soil quality, (ii) diversification of household energy sources can create possibilities for effective application of other strategies, and (iii) sequencing of interventions can improve effectiveness of conservation tillage. Our model-based approach demonstrates the interdependence of economic and ecological dynamics which preclude  blanket solution for poverty alleviation. Stylized models as developed here can  be used for testing effectiveness of different strategies given biophysical and economic settings in the target region.
\end{abstract}

\maketitle


\section{Introduction}

How to alleviate global poverty and eradicate hunger in places with low agricultural productivity are among humanity's greatest challenges. The concept of poverty traps as situations characterized by persistent, undesirable and reinforcing dynamics \citep{Haider} is increasingly being used to understand the relationship between persistent poverty and environmental sustainability \citep{Barrett15,BarrettCo,Lade}. How poverty and environmental degradation are conceptualized and represented in models can inform development interventions and thereby influence the effectiveness of those interventions \citep{Lade}. Previous poverty trap models have focused on environmental quality or pollution \citep{BarroSala,Smulders,Xepapadeas}, neglecting social-ecological interactions; have illustrated how positive feedback between wealth and technology can increase inequality and result in poverty traps through resource degradation \citep{Mirza}; have investigated relations between human health and poverty \citep{ng}; have used one-dimensional models that can lead to simplified conclusions and inappropriate policy outcomes \citep{Kraay}; have been static models  that cannot capture dynamic phenomena such as traps and feedbacks \citep{Barrett15}; or have been highly abstracted \citep{Lade}. 

\begin{figure}[ht]
    \centering
    \includegraphics[width=\linewidth]{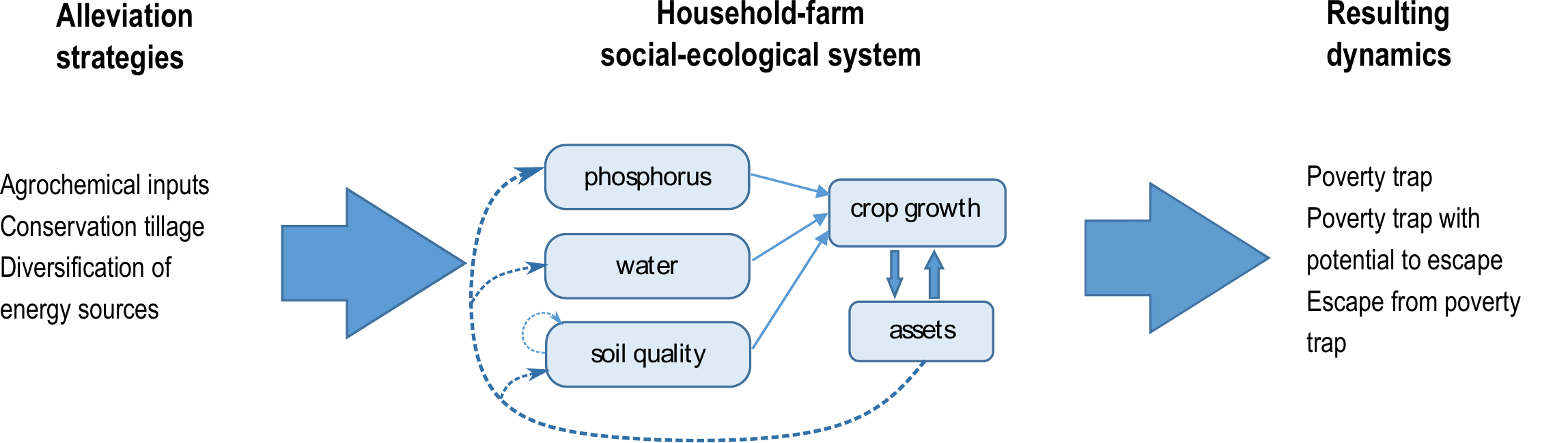}
    \caption{We investigate how phosphorus, soil quality and water interact with crop production and household assets. We treat a model of this household-farm social ecological system (middle section) with different combinations of interventions (left section) and observe the resulting poverty trap dynamics (right section). Some interventions involve households investing assets to improve phosphorus, soil quality and/or water levels (dashed line). In the model, soil quality can self-regenerate to a limited extent but phosphorus and water are reliant on continual replenishment.}
    \label{Figure1}
\end{figure}

Biophysical complexity is not often considered in poverty trap models and relations between agricultural interventions and social-ecological poverty trap dynamics remain unexplored. Partially because of this, development efforts tend to focus on blanket solutions, such as the ‘big push’: promoting external asset inputs, while neglecting a multitude of other factors affecting poverty.  \cite{Lade} highlighted the importance of linking economic, natural and human factors in explaining poverty traps and concluded that the usefulness of interventions depends on context, particularly the relationship between poverty and environmental degradation. We build on this study as a conceptual framework to address  knowledge gaps regarding the interplay between poverty and the biophysical environment in three ways: (1) we explore how biophysical complexity of the household-farm social-ecological system influences the dynamics of poverty traps in agroecosystems, (2) we assess the impact of development interventions on the dynamics of the system, and (3) we test the effectiveness of interventions (Figure \ref{Figure1}). To this end we have developed a  series of dynamical systems models that we use to test  diverse sequences of interventions for alleviating poverty. 

We describe biophysical complexity through factors that affect crop growth and limit food production  \citep{Drechsel,Rockstrom2000}, such as nutrients, especially phosphorus, water and soil quality. First, phosphorus is thought to have crossed a threshold of overuse at the global scale, leading to environmental consequences such as eutrophication \citep{Rockstrom09}, acidification \citep{Guo} and introduction of environmentally persistent chemicals or harmful elements in soil \citep{Carvalho,Pizzol,Roberts,Schnug}. However, at a local level many of the world’s poorest areas (e.g. Sub-Saharan Africa) suffer from a lack of soil nutrients, of which phosphorus is one of the main limiting factors for food production \citep{Nz,Verde}. Research indicates that global demand for phosphorus will rise over the remainder of the 21st century. At the same time, supply of high quality and accessible phosphate rock is likely to peak within the next few decades leading to increases in prices and decreases in affordability, mostly for low income countries \citep{Cordell}. Phosphorus application therefore presents a ‘double-edged sword’: in some cases it is necessary to overcome extreme levels of poverty and soil nutrient deficiency i.e. to break a poverty trap \citep{Lade}, but in other cases over application of fertilizers can have severe negative environmental consequences.
 
A second critical factor for crop growth is water. Rainfed agriculture plays a dominant role in food production, particularly in some of the poorest areas of the world, such as sub-Saharan Africa. Yield gaps are large and often caused by rainfall variability in occurrence and amount rather than by the total lack of water \citep{Rockstrom2000}. Because of this, investing in rainwater harvesting, water management and conservation practices, such as conservation tillage, is an important strategy for increasing food security and improving livelihoods. In small-scale semi-arid rainfed farming, these practices prove to be useful to mitigate drought and dry spells \citep{Rockstrom2003} or to allow diversification and cultivating high-value crops, which can be an important poverty alleviation strategy \citep{Burney}.
 
A third critical factor for crop growth is soil quality. It reflects complex interactions between soil physical, chemical and biological properties including environmental quality and soil’s contributions to health, food production and food quality. Including it in models bring additional level of realism and might explain human-environment relations \citep{Altieri,Bunemann,Parr,Verhulst,Thrupp}.

Agricultural interventions are a common strategy for poverty alleviation in developing countries. The interventions we consider here are largely carried out by actors external to the local community, such as non-governmental organisations (NGOs) or government programmes. For example, in the quest for an ‘African Green Revolution’ interventions to increase crop yields have been driven by: major cross-continental initiatives (Alliance for a Green Revolution in Africa), Millennium Villages Programmes (third party funded), donors (U.S. government’s Feed the Future program), and national governments with NGO’s implementing programmes at a local scale (Scoones and Thompson, 2011). In our models we focus on the implementation level of agricultural interventions.  

Inputs of fertilizers or improved seeds in the form of agricultural intensification schemes, or conservation tillage and use of manure as a fertilizer, while diversifying household energy sources are commonly used interventions. An intervention may influence one or more of the factors (assets, phosphorus, water or soil quality), thus ultimately influencing the dynamics of the whole agroecosystem. Since there are several factors at play, poverty alleviation  might require more than one intervention to be effective.

The aim of this paper is to develop a series of models that represent interlinked dynamics of assets, phosphorus, water and soil quality  and allow investigating their effects on the low-productivity poverty trap of many sub-Saharan communities \cite{Barrett06,Barrett08,Tittonell}. Furthermore, we use models to assess the effectiveness of different development interventions for various household-farm initial conditions. We begin by constructing a dynamical system model of an agroecosystem prior to any agricultural intervention and continue by developing three models representing changes in the dynamics of the agroecosystem due to agricultural interventions. Model assumptions are based on empirical evidence from the literature on nutrients, soil quality, water and economic aspects of poverty in arid areas as well as expert interviews (Table \ref{Table1}). 
We first analyse the baseline model without interventions and then sequentially assess the effectiveness of different alleviation strategies and their combinations (see Table 2 for a summary of the results and insights).
We conclude by discussing  our results and insights in relation to other theoretical and empirical work, and their importance for development practice and future research.

\section{The poverty trap models}

We use systems of nonlinear ordinary differential equations to set up a series of  multidimensional dynamical systems model of poverty traps. We begin by setting up a model which describes a household-farm system prior to any intervention and continue by presenting models incorporating different agricultural interventions. Table 1 contains our main assumptions about important factors for food production and the relationships between them derived from an extensive literature review and expert interviews. We use empirical evidence about poverty and agricultural production in arid regions, particularly Sub Saharan Africa, to extend a one dimensional theoretical poverty traps model towards a multi-dimensional and more realistic model.   

\begin{table}[h]\footnotesize
\begin{tabularx}{\textwidth}{m{5cm} m{11cm}}
\toprule
\bf{Model}    &  \bf{Assumptions and literature}  \\\midrule
The baseline model   & Rainfed agriculture \citep{Akhtar,Rockstrom2000,Rockstrom2003,Rockstrom2010}\newline
Manure used for household energy \citep{Int,Mekonnen,Niguisse} \newline
Agrochemicals (artificial fertilizers) are not used \citep{Druilhe} \newline
Water, phosphorus and assets are necessary for crop production \citep{Kataria}
\\\midrule
Scenario 1: \newline Input of agrochemicals & 1a: Endogenous strategy (agrochemicals purchased with savings) \newline
Agrochemicals increase phosphorus level in the soil, but may have negative effect on soil quality. \citep{Guo,Loreau,Pizzol,Roberts,Schnug} \newline
Soil quality can regenerate. \citep{Bunemann,Smulders,Xepapadeas} \newline
Improved water conditions are enabled by rainwater harvest. \citep{Enfors08,Enfors13,Yosef} 
\newline 
1b: Exogenous strategy (purchasing through external support or loan) \newline
Strong negative effect of agrochemicals on soil quality \citep{Geiger,Pizzol,Roberts,Savci,Schnug} 
 \\\midrule
Scenario 2: \newline Diversification of household \newline energy sources  & Different household energy sources in SSA. Diverse energy sources allows manure to be used as fertiliser instead of fuel. \citep{Int} \newline
Manure improves soil quality and nutrient level. \citep{Bationo,DeAngelis,Kihanda,McConville,Pretty,Probert,Wanjekeche} 
\newline 
Improved water conditions are enabled by rainwater harvest. \citep{Enfors13,Yosef}
 \\\midrule
Scenario 3: \newline Conservation tillage  & 3a: Conservation tillage with phosphorus as limiting factor and no additional nutrient input \citep{DeAngelis,Ito,McConville,Pretty,Verhulst}
\newline
3b: Conservation tillage with phosphorus as limiting factor and artificial fertilizer/manure application \citep{DeAngelis,Ito,McConville,Pretty,Verhulst,Wanjekeche}
\newline
3c: Conservation tillage with water as limiting factor \citep{Asmamw}
 \\\bottomrule
\end{tabularx}
\smallskip
\caption{List of models and underlying assumptions.}\label{Table1}
\end{table}

These assumptions enable us to construct causal loop diagrams (Figures 2-4) and to choose state variables and functional forms for our dynamical systems. The key assumptions are:

\begin{enumerate}
\item Phosphorus content of soils. Agricultural production removes phosphorus from crop producing soils, which if not balanced by agroecological methods \citep{Altieri}, or application of organic or artificial fertilizers limits crop growth and leads to lower yields \citep{Drechsel}. 

\item Water content of soils. Although rainfed agriculture is a widespread practice, it cannot always provide optimal water conditions, especially under the conditions of climate change. 

\item Soil quality. Soil quality is a more complex variable than the nutrient content of soils or its capacity to produce crops alone. Accordingly, we model its dynamics separately to that of phosphorus and acknowledge that it might be self-regenerating. 

\item Assets. Assets such as agrochemicals, improved seeds, and tools used for agriculture, supports agricultural production and can be a limiting factor for people living below the poverty line \citep{Druilhe,Kataria}. We extend standard neoclassical dynamics of assets in which profit can be consumed or saved for investment in future production. Specifically, we implement a ‘savings trap’, in which households have a lower savings rate at low asset levels, leading to a trap in which they are unable to accumulate enough assets to escape poverty \citep{Kraay}.
\end{enumerate}

While each of these variables has its own dynamics, they also interact in complex ways (Figure 2A). Understanding the resulting dynamics is important for designing effective poverty alleviation strategies. We use the household scale because we seek to investigate the consequences of household-level decision making and because agricultural interventions often focus on smallholder farms \citep{Nz,Probert,Rockstrom2000,Verde}.  

We analyse dynamics of the system by studying its attractors and basins of attraction. An attractor is a state (or set of states) to which the system tends over time starting from an initial state. It is defined by values of state variables, e.g. assets, phosphorus, water or soil quality. A basin of attraction is a set of all states of the system which tend over time towards the same attractor.


\subsection{The baseline model}

The purpose of the baseline model is to describe dynamics of a typical low-income farming household in sub-Saharan Africa. Due to lack of assets and external inputs, artificial fertilizers are not used. Manure is used as a household energy source and rain is the only source of water. Key factors for food production are assets, phosphorus, water and soil quality and here we briefly explain their role as state variables in the models.

\underline{Assets.} The neoclassical economic theory of growth defines production output $y$ as a function $f(k)$, where $k$ is capital. In the economic literature, it is common to consider different forms of physical capital, such as infrastructure or machinery, but here we include per capita assets $k_a$, phosphorus $k_p$, water $k_w$ and soil quality $k_q$. Like previous works on poverty traps \citep{Lade,Kraay} we use a Solow model \citep{BarroSala} to model asset dynamics,
\begin{align}\label{solow}
    \frac{dk_a}{dt} &= s(k_a)f(k_a,k_p,k_w,k_q)-(\delta_a+r)k_a,
\end{align}
where $s(k_a)$ is a nonlinear savings rate \citep{Kraay} and $f(k_a,k_p,k_w,k_q)$ is a production function with assets, phosphorus, water and soil quality as necessary variables.  In other words, the value of the production function $f$ is zero given zero assets, phosphorus, water or soil quality, making crop production impossible in those cases. We assume that the function $f$ is the Cobb-Douglas production function of the form
\begin{align}\label{cobb}
    f(k_a,k_p,k_w,k_q)=Ak_a^{\alpha_a}k_p^{\alpha_p}k_w^{\alpha_w}k_q^{\alpha_q}, \quad A>0, \quad \alpha_a+\alpha_p+\alpha_w+\alpha_q \le 1,
\end{align}
or some of its simpler variants
\begin{align}\label{cobb1}
    f(k_a,k_p,k_w)=Ak_a^{\alpha_a}k_p^{\alpha_p}k_w^{\alpha_w}, \quad A>0, \quad \alpha_a+\alpha_p+\alpha_w \le 1,
\end{align}
or
\begin{align}\label{cobb2}
    f(k_a,k_p,k_q)=Ak_a^{\alpha_a}k_p^{\alpha_p}k_q^{\alpha_q}, \quad A>0, \quad \alpha_a+\alpha_p+\alpha_q \le 1,
\end{align}
where $A$ is a constant productivity term.  Parameters $\delta_a$ and $r$ in equation (\ref{solow}) denote assets depreciation rate and population growth rate, respectively. Both of them affect assets growth rate negatively. For more details on derivation of equation (\ref{solow}) we refer readers to Appendix A.

An s-shaped function for the savings rate $s(k_a)$ allows formation of savings traps \citep{Kraay}. We assume that it has the following form
\begin{align}\label{savings}
    s(k_a)=\frac{s_1}{1+e^{-s_2k_a+s_3}}, \quad s_1,s_2>0,s_3\ge0.
\end{align}

\underline{Phosphorus.} Phosphorus cycling in an agroecosystem system begins with phosphorus in the soil. From there, it is taken up by plants and transported though the food web to the consumers on or off-farm. By-products of food production, manure or human waste are usually not recycled and used as  fertilizers (Table \ref{Table1}). Even if there is no agricultural production, poor people may rely on collecting biomass to secure their livelihood. In addition to this, short intensive rainfalls which occur more frequently contribute to phosphorus loss by washing away top soil layers. Thus, the amount of phosphorus in the soil needed for crop growth is constantly declining, which we describe by the following equation:
    \begin{align}\label{p0}
        \frac{dk_p}{dt}= -\delta_pk_p, \quad \delta_p>0,
    \end{align}
where $\delta_p$ is the phosphorus loss rate.

Phosphorus loss can depend on assets levels (increasing with assets to reflect consequences of intensified production), in which case the phosphorus loss rate can be written as $\delta_p(k_a)=\delta_p(1+\frac{d_1k_a}{d_2+k_a})$, where $d_1\ge 0$ and $d_2>0$. Having this more complicated phosphorus loss rate will not affect qualitative behavior of the model since the term in the brackets is always positive and the only solution to equation $\frac{dk_p}{dt}=0$ is $k_p=0$. Because of this we formulate our models using the assets independent loss rate as in equation (\ref{p0}).
    
\underline{Water.} We assume that rain is the only water supply. A portion of rain water is used by plants for their growth, while the rest is lost due to evaporation, leaking or sinking into lower soil layers inaccessible to plants. Therefore, water dynamics satisfies the following equation:
    \begin{align}\label{w0}
        \frac{dk_w}{dt} = r_w-\delta_wk_w, \quad r_w\ge 0, \, \delta_w>0,
    \end{align}
where $r_w$ is the amount of water gained by rainfall and $\delta_w$ is the water loss rate. 
    
\underline{Soil quality.} Soil quality refers to the soil's properties that enable food production, such as soil structure, amount of pollutants or microorganisms, but exclude soil's nutrient content since it is modeled through phosphorus and water. The purpose of having this variable is to introduce biochemical and biophysical complexity of soil into models and to enable modelling various influences human actions may have. Since soil quality can be related to populations of organisms that live in or on soil or contribute to soil organic matter when they decompose, we assume that soil quality can regenerate (Table \ref{Table1}) following logistic growth:
\begin{align}\label{q0}
    \frac{dk_q}{dt} = r_qk_q\left(1-\frac{k_q}{Q}\right),  \quad r_q\ge 0, \, Q>0,
\end{align}
where $r_q$ is soil's quality recovery rate and $Q$ its carrying capacity. If soil quality represent soil's capacity to absorb pollution, then its values vary  between zero and some positive upper bound $Q$ and the ecological processes that give this ability can be modeled using the logistic model.

\begin{figure}[ht]
\begin{minipage}[b]{0.35\textwidth} 
\begin{subfigure}[b]{\textwidth}\resizebox{\linewidth}{!}{
\begin{tikzpicture}[->,>=stealth',auto,node distance=3cm,
  thick,main node/.style={rectangle, rounded corners, minimum height=2em, draw,font=\bfseries}]
\node[main node] (1) {CG};
\node[main node] (2) [right of=1] {$k_a$};
\node[main node] (4) [left of=1,xshift=1cm] {$k_p$};
\node[main node] (5) [below of=1,yshift=1.5cm] {$k_w$};
\node[main node] (7) [left of=4,xshift=1cm] {M};
\node[main node] (8) [below of=7,yshift=1.5cm] {HES};
\draw [->] (1.north) to [out=60,in=120] node[pos=0.5,above]{+}([xshift=-0.2cm]2.north) ;
\draw [->] ([xshift=-0.2cm]2.south) to [out=240,in=300] node[pos=0.5,below]{+} ([xshift=0.2cm]1.south);
\draw [->] (5.north) to  node[pos=0.5,left]{+} (1.south);
\draw [->,dashed] (7.east) to node[pos=0.5,above]{+} (4.west);
\draw [->] (7.south) to node[pos=0.5,left]{+} (8.north);
\draw [->] (4.east) to node[pos=0.5,above]{+} (1.west);
\end{tikzpicture}}
\caption{}
\end{subfigure}
\end{minipage}
\hspace*{0.5cm}
\begin{minipage}[b]{0.35\textwidth}
\begin{subfigure}[b]{\linewidth}
\includegraphics[width=0.8\linewidth]{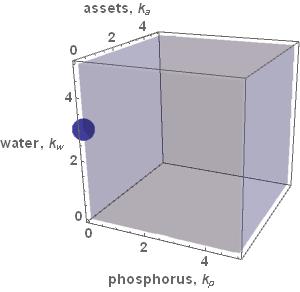}
\caption{}
\end{subfigure}
\end{minipage}
\caption{Causal loop diagram (A) and state space plot (B) for system (\ref{baseline1}) before interventions. Abbreviations denote: M manure, HES household energy source, CG crop growth, $k_a$ assets, $k_p$ phosphorus, $k_w$ water. The dashed line in figure (A) indicates that insignificant, if any, amount of manure is used as a fertilizer. The blue disc in figure (B) represents a unique attractor of this system, whose basin of attraction is the whole state space. The parameters are $s_1=0.1, s_2=10, s_3=20, A=10, \alpha_a=0.3, \alpha_p=0.3, \alpha_w=0.3, \delta_a=1, \delta_p=1, r_w=1.5, \delta_w=0.5$. }
\label{Figure2}
\end{figure}
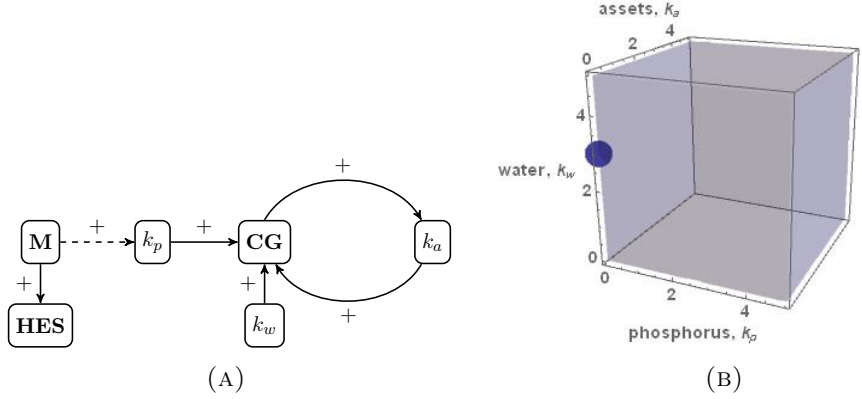

The baseline scenario is represented by the causal loop diagram in Figure \ref{Figure2} and the corresponding dynamical system
\begin{equation}\label{baseline1}
    \begin{aligned}
     \frac{dk_a}{dt} &= s(k_a)f(k_a,k_p,k_w)-(\delta_a+r)k_a, \\
     \frac{dk_p}{dt} &= -\delta_pk_p, \\
     \frac{dk_w}{dt} &= r_w-\delta_wk_w,
    \end{aligned}
\end{equation}
where we used equations (\ref{solow}), (\ref{p0}) and (\ref{w0}) to describe dynamics of each variable and the function $f$ is given by (\ref{cobb1}).

In what follows, we will present three scenarios which describe common agricultural interventions. Applications of agrochemicals, including artificial fertilisers, preserves the openness of the agroecosystem \citep{DeAngelis}. Two other interventions, conservation tillage and household energy diversification, lead to a more closed agroecosystem in which energy and matter is recycled internally. These interventions are sometimes accompanied by water preserving techniques and we include them in our models to show effects of different water regimes.


\subsection{Scenario 1: Input of agrochemicals}

\begin{figure}[ht]
\begin{minipage}[b]{\textwidth}
\begin{minipage}[b]{0.3\textwidth} 
\begin{subfigure}[b]{\textwidth}\resizebox{\linewidth}{!}{
\begin{tikzpicture}[->,>=stealth',auto,node distance=3cm,
  thick,main node/.style={rectangle, rounded corners, minimum height=2em, draw,font=\bfseries}]
\node[main node] (1) {CG};
\node[main node] (2) [right of=1] {$k_a$};
\node[main node] (3) [left of=1,yshift=-1cm] {$k_q$};
\node[main node] (4) [left of=1,yshift=1cm] {$k_p$};
\node[main node] (5) [below of=1,yshift=1cm] {$k_w$};
\node[main node] (6) [left of=4,yshift=-1cm] {AF};
\node[main node] (7) [left of=5,xshift=-2cm] {M};
\node[main node] (8) [below of=7,yshift=1.5cm] {HES};
\node[main node] (10) [below of=5, yshift=1.5cm] {WM};

\draw [->] (1.north) to [out=60,in=120] node[pos=0.5,above]{+}([xshift=-0.2cm]2.north) ;
\draw [->] ([xshift=-0.2cm]2.south) to [out=240,in=300] node[pos=0.5,below]{+} ([xshift=0.2cm]1.south);
\draw [->] (5.north) to  node[pos=0.5,left]{+} (1.south);
\draw [->] ([yshift=0.1cm]6.east) to node[pos=0.5,above]{+} ([yshift=0.1cm]4.west);
\draw [->] ([yshift=-0.1cm]6.east) to node[pos=0.5,above]{-} ([yshift=-0.1cm]3.west);
\draw [->] (7.south) to node[pos=0.5,left]{+} (8.north);
\draw [->] (4.east) to node[pos=0.5,above]{+} ([yshift=0.1cm]1.west);
\draw [->] (3.east) to node[pos=0.5,below]{+} ([yshift=-0.1cm]1.west);
\draw [->] (10.north) to node[pos=0.5,left]{+} (5.south);
\draw [->] ([xshift=-0.3cm]3.south) arc (130:410:4mm) node [pos=0.5, below] {+};
\draw [dgreen,->] (2.north) to [out=110,in=70] node[pos=0.5,below]{+} (6.north);
\draw [blue,->] (2.south) to [ out=270,in=0] node[pos=0.5,left]{+} (10.east);
\end{tikzpicture}}
\caption{}
\end{subfigure}
\end{minipage}
\hspace*{1cm}
\begin{minipage}[b]{0.3\textwidth}
\begin{subfigure}[b]{\linewidth}
\includegraphics[width=\linewidth]{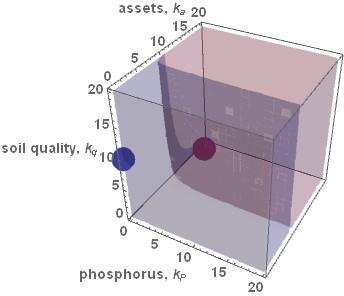}
\caption{}
\end{subfigure}
\end{minipage}
\hspace*{1cm}
\begin{minipage}[b]{0.3\textwidth}
\begin{subfigure}[b]{\linewidth}
\includegraphics[width=\linewidth]{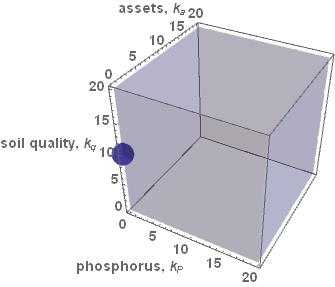}
\caption{}
\end{subfigure}
\end{minipage}
\end{minipage}
\vspace*{5mm}
\begin{minipage}[b]{\textwidth} 
\begin{minipage}[b]{0.3\textwidth} 
\begin{subfigure}[b]{\textwidth}\resizebox{\linewidth}{!}{
\begin{tikzpicture}[->,>=stealth',auto,node distance=3cm,
  thick,main node/.style={rectangle, rounded corners, minimum height=2em, draw,font=\bfseries}]
\node[main node] (1) {CG};
\node[main node] (2) [right of=1] {$k_a$};
\node[main node] (4) [left of=1,xshift=1cm] {$k_p$};
\node[main node] (5) [below of=1,yshift=1cm] {$k_w$};
\node[main node] (7) [left of=4,xshift=1cm] {M};
\node[main node] (9) [below of=7, yshift=-0.5cm] {HES};
\node[main node] (10) [below of=5, yshift=1.5cm] {WM};

\draw [->] (1.north) to [out=60,in=120] node[pos=0.5,above]{+}([xshift=-0.2cm]2.north) ;
\draw [->] ([xshift=-0.2cm]2.south) to [out=240,in=300] node[pos=0.5,below]{+} ([xshift=0.2cm]1.south);
\draw [->] (5.north) to  node[pos=0.5,left]{+} (1.south);
\draw [->,thick] (7.east) to node[pos=0.5,above]{+} (4.west);
\draw [->] (4.east) to node[pos=0.5,above]{+} (1.west);
\draw [->,blue] (10.north) to node[pos=0.5,left]{+} (5.south);
\draw [dgreen,->] (2.east) to [ out=300,in=330] node[pos=0.5,below]{+} (9.south);
\draw [blue,->] (2.south) to [ out=300,in=0] node[pos=0.5,left]{+} (10.east);
\end{tikzpicture}}
\caption{}
\end{subfigure}
\end{minipage}
\hspace*{1cm}
\begin{minipage}[b]{0.3\textwidth}
\begin{subfigure}[b]{\linewidth}
\includegraphics[width=\linewidth]{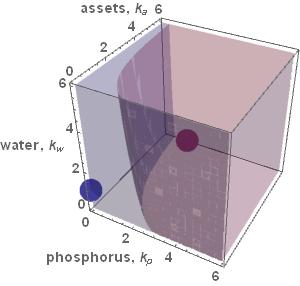}
\caption{}
\end{subfigure}
\end{minipage}
\hspace*{1cm}
\begin{minipage}[b]{0.3\textwidth}
\begin{subfigure}[b]{\linewidth}
\includegraphics[width=\linewidth]{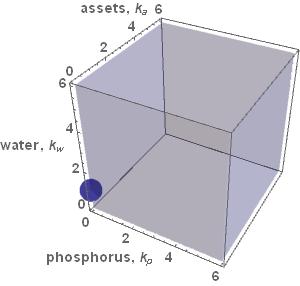}
\caption{}
\end{subfigure}
\end{minipage}
\end{minipage}
\caption{Causal loop diagram and state space with attractors and basins of attraction for agrochemical use (A-C) and diversification of household energy sources (D-F). Abbreviations denote: M manure, HES household energy source, AF combination of artificial fertilizer, improved seeds and chemicals, CG crop growth, WM water management, $k_a$ assets, $k_p$ phosphorus, $k_w$ water and $k_q$ soil quality. The blue and green lines in (A) represent endogenous water management and agrochemicals input.  The green line in (D) is for endogenous household energy sources. The blue and purple discs represent attractors and colored volumes are corresponding basins of attraction. (B) Good water conditions and mild negative effect of agrochemicals on soil quality; system (\ref{af}) with $s_1=0.25, s_2=2.5, s_3=20, A=10, \alpha_a=0.4, \alpha_p=0.3, \alpha_q=0.2, \delta_a=0.7, c_1=1, c_2=20, c_3=1, c_4=4, \delta_p=0.2, r_q=1, Q=10$. (C) Good water conditions and strong negative effect of agrochemicals on soil quality; $c_3=4$. (E) Sufficient amount of nutrient rich manure and good water conditions; system (\ref{energy}) with $s_1=0.1, s_2=1, s_3=0, A=10, \alpha_a=0.3, \alpha_p=0.3, \alpha_q=0.2, \delta_a=0.5, c_1=1, c_2=5, c_3=1, c_4=1.8, \delta_p=0.2, r_w=1, c_5=1, c_6=40, \delta_w=1$. (F) Insufficient amount or nutrient poor manure and good water conditions; $c_1=0.5$.}
\label{Figure3}
\end{figure}
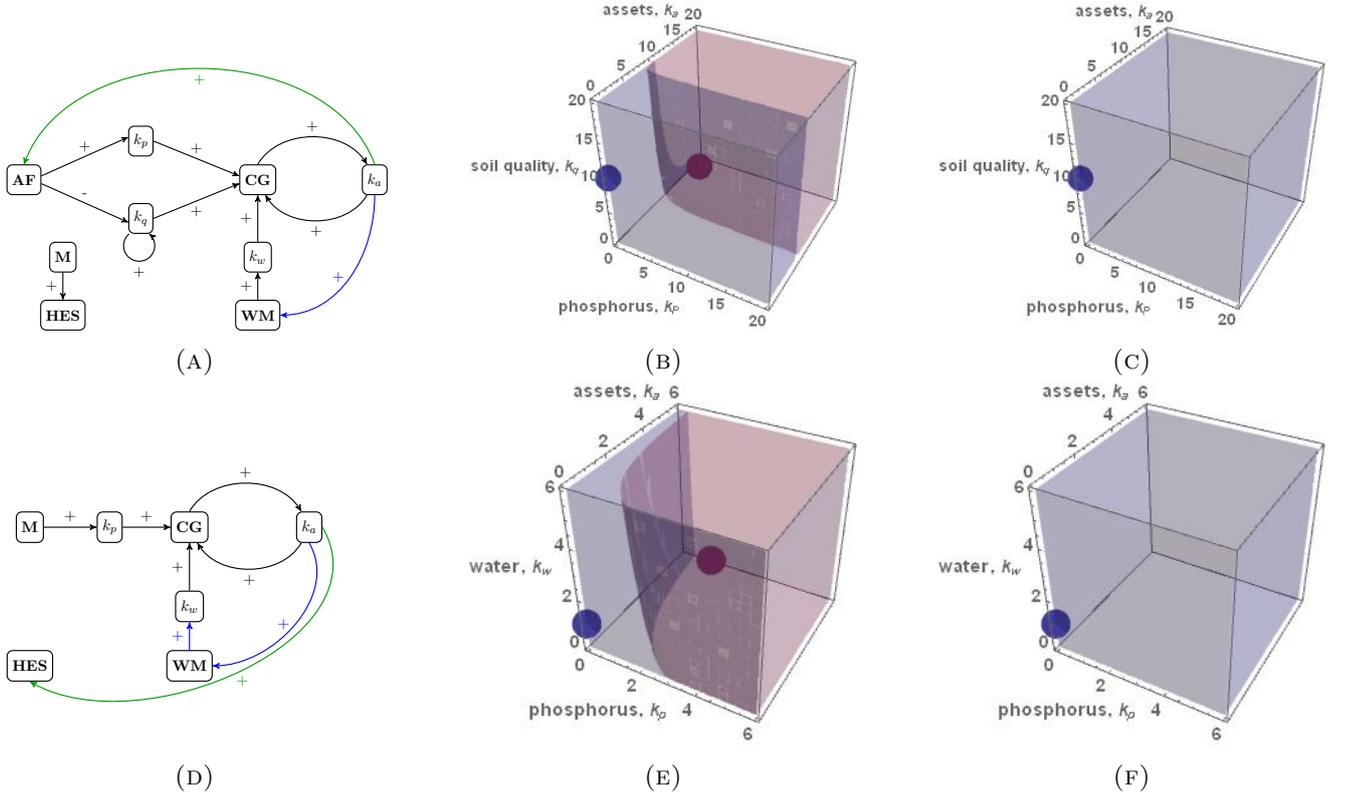

Using improved seeds is usually accompanied by application of combinations of agrochemicals, such as fertilizers, herbicides and pesticides. Apart from the intended effects of increasing phosphorus levels in the soil, side effects such as soil acidification and loss of biodiversity have been observed  (Table \ref{Table1}). In order to describe this dual effects of agrochemicals and study corresponding dynamics, we use assets, phosphorus and soil quality as state variables for the system. We model that household invests part of its assets in agrochemicals. We also assume that the household invests a part of its income into water management and because of this water is not a limiting factor for crop growth. 
The causal loop diagram is given in Figure \ref{Figure3}A and mathematical formulation of the model is obtained by modifying model (\ref{baseline1}) and reads as follows:
\begin{equation}\label{af}
    \begin{aligned}
     \frac{dk_a}{dt} &= s(k_a)f(k_a,k_p,k_q)-(\delta_a+r)k_a, \\
     \frac{dk_p}{dt} &= I_p(k_a)-\delta_pk_p, \\
     \frac{dk_q}{dt} &= r_qk_q\left(1-\frac{k_q}{Q}\right)-I_q(k_a)k_q,
    \end{aligned}
\end{equation}
where $f$ is defined by (\ref{cobb2}), $I_p(k_a)$ is the increase in phosphorus due to artificial fertilizer and $I_q(k_a)$ is the negative effect of agrochemicals on soil quality. Positive contributions of fertilizer to soil's phosphorus content are limited and the same is true for negative effects of agrochemicals on soil quality. We assume that these functions have the form
$$I_p(k_a)=\frac{c_1k_a^2}{c_2+k_a^2} \quad\mbox{and}\quad I_q(k_a)=\frac{c_3k_a}{c_4+k_a}, \quad c_1,c_2,c_3,c_4>0.$$

The coupled system (\ref{af}) incorporates the positive feedback between assets and phosphorus and the negative feedback between assets and soil quality. The parameters $c_1$ and $c_2$ define positive contributions of fertilizer to soil's phosphorus content, and the parameters $c_3$ and $c_4$ define strength of negative effect of agrochemicals on soil quality. Depending on the parameter values, the system can have different number of attractors. 

\subsection{Scenario 2: Diversification of household energy sources}

According to the assumptions in Table \ref{Table1}, manure is a valuable fertilizer, but most of it is used as a household energy source. We model a situation when a household invests some of its assets into new energy source and more fuel efficient technologies and uses manure as a fertilizer. We also model that farmers invest part of their assets in rainwater harvesting technologies, which improves water conditions. This leads us to the causal loop diagram in Figure 3D and following dynamical system:
\begin{equation}\label{energy}
    \begin{aligned}
     \frac{dk_a}{dt} &= s(k_a)f(k_a,k_p,k_w)-(\delta_a+r)k_a, \\
     \frac{dk_p}{dt} &= I_p(k_a,k_p)-\delta_pk_p, \\
     \frac{dk_w}{dt} &= r_w+I_w(k_a)k_w-\delta_wk_w,
    \end{aligned}
\end{equation}
where $f$ is given by (\ref{cobb1}) and functions $I_p(k_a,k_p)$ and $I_w(k_a)$ have the form
$$I_p(k_a,k_p) = \frac{c_1k_a^2}{c_2+k_a^2}\cdot\frac{c_3k_p}{c_4+k_p} \quad\mbox{and}\quad I_w(k_a)=\frac{c_5k_a^2}{c_6+k_a^2}, \quad c_i>0, i=\overline{1,6}.$$
The first factor in the function $I_p(k_a,k_p)$ is related to the amount of manure that can be gained by energy source diversification. We choose a s-shaped function of assets since the available manure is limited. The second factor in the function $I_p(k_a,k_p)$ is related to manure quality, which is measured by the amount of phosphorus manure contains and this content depends on the environment (low in degraded environment, high in good environment). Water gains are modelled using function $I_w(k_a)$.


\subsection{Scenario 3: Conservation tillage}

Conservation tillage is a method which includes a range of tillage practices aimed to increase water infiltration and nutrient conservation and decrease water and nutrient loss through evaporation, leaching and erosion \citep{Busari}. Since conservation tillage does not provide additional nutrient or water input, we model it using the baseline model (\ref{baseline1}) and Figure \ref{Figure2} with reduced phosphorus and water loss rates. Depending on its effectiveness, conservation tillage can reduce or even eliminate phosphorus or water loss. The outcome of the intervention in the first case is still depletion of phosphorus, but at a slower pace than in the baseline model. 

If tillage eliminates phosphorus loss, $\frac{dk_p}{dt}=0$, the corresponding dynamical system is then
\begin{equation}\label{ct}
    \begin{aligned}
     \frac{dk_a}{dt} &= s(k_a)f(k_a,k_w)-(\delta_a+r)k_a, \\
     \frac{dk_w}{dt} &= r_w-\delta_wk_w,
    \end{aligned}
\end{equation}
where $f(k_a,k_w)=Ak_a^{\alpha_a}k_w^{\alpha_w}$ and the productivity term $A$ incorporates effects of constant phosphorus level on crop growth.

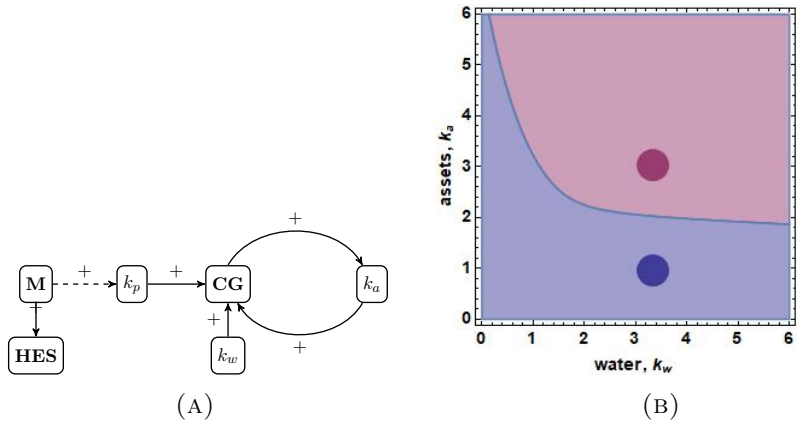
\begin{figure}[H]
\begin{minipage}[b]{0.3\textwidth} 
\begin{subfigure}[b]{\textwidth}\resizebox{\linewidth}{!}{
\begin{tikzpicture}[->,>=stealth',auto,node distance=3cm,
  thick,main node/.style={rectangle, rounded corners, minimum height=2em, draw,font=\bfseries}]
\node[main node] (1) {CG};
\node[main node] (2) [right of=1] {$k_a$};
\node[main node] (4) [left of=1,xshift=1cm] {$k_p$};
\node[main node] (5) [below of=1,yshift=1.5cm] {$k_w$};
\node[main node] (7) [left of=4,xshift=1cm] {M};
\node[main node] (8) [below of=7,yshift=1.5cm] {HES};
\draw [->] (1.north) to [out=60,in=120] node[pos=0.5,above]{+}([xshift=-0.2cm]2.north) ;
\draw [->] ([xshift=-0.2cm]2.south) to [out=240,in=300] node[pos=0.5,below]{+} ([xshift=0.2cm]1.south);
\draw [->] (5.north) to  node[pos=0.5,left]{+} (1.south);
\draw [->,dashed] (7.east) to node[pos=0.5,above]{+} (4.west);
\draw [->] (7.south) to node[pos=0.5,above]{+} (8.north);
\draw [->] (4.east) to node[pos=0.5,above]{+} (1.west);
\end{tikzpicture}}
\caption{}
\end{subfigure}
\end{minipage}
\hspace*{0.5cm}
\begin{minipage}[b]{0.35\textwidth}
\begin{subfigure}[b]{\linewidth}
\includegraphics[width=0.8\linewidth]{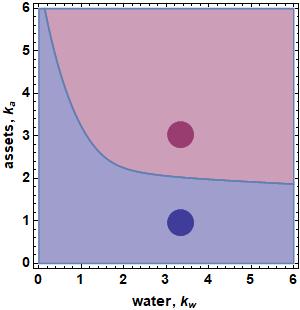}
\caption{}
\end{subfigure}
\end{minipage}
\caption{Causal loop diagram (A) and state space plot for conservation tillage which eliminates phosphorus loss (B) for high phosphorus content and good water conditions; system (\ref{ct}) with $s_1=0.1, s_2=10, s_3=20, A=6, \alpha_a=0.4, \alpha_w=0.4, \delta_a=1,  r_w=1, \delta_w=0.2$.}
\label{Figure4}
\end{figure}

\section{Results}
\subsection{Multi-dimensional poverty trap model of household agriculture (baseline model)}

The baseline model (system (\ref{baseline1}); Figure 2) represents agroecological dynamics at the household-farm scale in sub-Saharan Africa prior to any agricultural intervention. The soil gradually loses phosphorus, reducing crop growth and income and pushing household to a poor state characterized by a low asset level and phosphorus depletion (blue disc in Figure 2B). Regardless of the initial levels of water, nutrients and assets, the household-farm system always reaches the low well-being attractor due to losses of phosphorus and lack of replenishment. 

Short term external asset inputs provide only a change in the initial conditions, but leave the attractor unchanged, and because of this, they are unable to alleviate persistent poverty.
This result suggests that short-term external poverty alleviation interventions are not sufficient in this situation; some structural change in household-farm dynamics is required to enable the existence of a non-poor state. Disrupting the negative phosphorus balance is necessary and it can be achieved by external phosphorus inputs or by preventing its loses. 

In what follows we analyse Scenario 1, 2 and 3 and assess their poverty alleviation potential. 


\subsection{Agrochemical inputs can reinforce poverty and degradation of soil quality}

Using improved seeds and agrochemicals is a common practice in agricultural intensification, but there is evidence both of success \citep{Carvalho,Weight} and failure \citep{Fischer}. In order to visualize the outcome of agrochemical applications, we assume constant water levels and present the case where production is not water-limited (system (\ref{af}); Figure 3A-C). 

Cases when moderate agrochemical application \citep{Snapp} leads to mildly negative effects on soil quality (Table \ref{Table1}) is represented in Figure 3B.
In addition to the poor attractor (blue disc), an alternative high well-being attractor exists in our model (purple disc). At high asset levels farmers have the resources to purchase agrochemicals which increases farm productivity, although at the cost of soil quality. At low asset levels, farmers do not have the resources to increase productivity by purchasing agrochemicals and therefore the low asset level attractor remains in the model (blue disc). To escape the low asset level poverty trap, agrochemical application would therefore need to occur in conjunction with an external asset input.

Evidence is accumulating that agrochemical application can be severely harmful for soil quality (Table \ref{Table1}). This case is represented in Figure 3C and strong negative effect of agrochemicals is obtained by increasing $c_3$.  If the harmful effects of agrochemicals on productivity via degradation of soil quality outweigh improvements in productivity via improved soil nutrient levels, poverty is the likely outcome.

Our results show how the dual consequences, both positive and negative, of agrochemical inputs can in fact reinforce as well as alleviate poverty, depending on which effect is stronger in a specific situation. Application of agrochemicals can also fail since they need to be recurrently applied, which makes farmers dependent on government support \citep{Gerber} or forces farmers to prioritise investment in artificial fertilizers, though we do not model these mechanisms here. 


\subsection{Prioritising diversification of energy sources can establish the conditions for effective application of other strategies}

Application of manure to soils can improve crop yields by increasing nutrient content of the soil and improving soil quality \citep{Bationo,Kihanda,Probert}. However, most manure in Sub-Saharan Africa is used as a household energy source \citep{Mekonnen,Niguisse}, and little is left to be used as a fertilizer. Diversifying household energy sources using gas, charcoal or electricity allows manure to be used for soil fertilization \citep{Int} and investment in rainwater harvesting technologies improves water levels \citep{Yosef}. In practice, such investments in alternative energy and rainwater technologies may also require capacity building and behavioural change as well as access to an affordable supply of the alternative technology. We here provide insights gained from model (\ref{energy}). 

A case when there is sufficient quantity of nutrient rich manure is given in Figure 3E. Investment in energy diversification and rainwater harvesting can introduce an alternative attractor at higher phosphorus levels (purple disc). An attractor remains at low phosphorus levels (blue disc), where households remain trapped in poverty because they cannot afford sufficient spending to increase nutrient or water levels or because the soil is too degraded. To transition out of the poor and degraded attractor to the new attractor would require an initial input of external assets and increasing phosphorus levels in the soil prior to manure application. 

If insufficient quantities of manure is available, or its nutrient content is low, only the initial poor and degraded attractor remains (Figure 3F). In such cases, manure for fertilization is inadequate for obtaining higher yields and escaping poverty, as for example was observed in \citep{Wanjekeche}. A similar result is obtained if the total rainfall is too low. In these cases, water management cannot improve water conditions, leading to low production and even large inputs of external assets will not allow households to escape their poverty trap.

Our results show that energy diversification and rainwater harvesting can be prerequisites for effective application of other poverty alleviation strategies. In a highly degraded environment, it may be necessary to also follow diversification of energy sources with a combination of manure and artificial fertilisers \citep{Ito} to raise phosphorus content to a level needed for production.


\subsection{Sequencing of interventions matters for effectiveness of conservation tillage}

Conservation tillage is a popular intervention method in Sub-Saharan Africa \citep{Enfors11,FR}. It includes a range of tillage practices aimed to increase water infiltration and nutrient conservation and decrease water evaporation, nutrient leaching and soil erosion \citep{Busari}.  Conservation tillage does not provide any external inflows of nutrients or water, but it reduces their loss. We used system (\ref{baseline1}) and (\ref{ct}) to study how the effectiveness of conservation tillage, water conditions, phosphorus and asset levels affect households in poverty traps.

If conservation tillage reduces, but does not eliminate nutrient leaching, the system dynamics have the same properties and long-term behavior as the baseline model (system (\ref{baseline1}); Figure 2). Regardless of the initial conditions, a household will end in poverty, likely at a slower pace than in the baseline case. In this case, conservation tillage will need to be paired with continual application of manure or artificial fertiliser, as for example in \citep{Ito} (and which in our model leads to the same two-attractor configuration as in Figure 3B). Low water levels can also limit crop growth and the effectiveness of conservation tillage for any nutrient level. In this case, improving water levels through rainwater harvesting technologies such as small-scale water catchments \citep{Enfors13} would prepare the conditions for conservation tillage to be useful. 

If tillage eliminates (or almost eliminates) nutrient leaching, phosphorus levels will be conserved over time. System (\ref{ct}) has two attractors for sufficiently high phosphorus and water levels
(Figure \ref{Figure4}). Higher levels of phosphorus that enable productivity can introduce an alternative attractor with higher asset level. Because the low asset level attractor remains after conservation tillage, additional external asset inputs may be required along with or after conservation tillage to allow the household to escape the poverty trap.

If the phosphorus (or water) level is low, the subsequent low levels of agricultural production keep the household in poverty for any levels of assets and water (or phosphorus) and only one attractor exists (Figure 5 in Supplementary Information). 

The dynamic nature of our model shows how the sequence of interventions can critically affect whether conservation tillage can allow a household to escape a poverty trap. A sequence of interventions starting with nutrient application, then conservation tillage accompanied by a one-off external asset input may be most effective, especially when initial nutrient levels are low. If conservation tillage does not eliminate nutrient leaching, farmers may need to invest in energy diversification or artificial fertilisers to allow continued application of nutrients.

\subsection{Case study example}

Our models are not designed to represent a particular real-world case study. They aim to capture key dynamics and contextual factors found in the context of rural poverty in a stylized way. Simplicity of our models allows testing and assessing the consequences of a combination of factors assumed to be present in a specific case before designing an intervention or building an empirical model. Their main purpose is thus to support a process of thinking through complex interactions that are difficult or impossible to assess in an empirical study. We demonstrate the value-added of using dynamical systems modelling as a thinking tool to support development interventions in agricultural contexts through a case study. In North-Eastern Tanzania, \citet{Enfors13} conducted a study on how water management technology would influence agro-ecosystem dynamics. The study outlines alternative development trajectories based on specific social-ecological feedbacks and the role of small-scale water systems in breaking trap dynamics. 

Our modelling approach could help an implementing body (an NGO for example) that aims to introduce conservation tillage (as a water saving intervention) to compare possible outcomes depending on the households' initial conditions and local biophysical and economic context. For example, conservation tillage can preserve nutrients and water but it will only be effective if there are enough nutrients and water in the soil as a starting point (Scenario 3, Figure 4, Figure 5 in Appendix A). Conservation tillage should be complemented with water management to increase the level of water in case of severe drought, demonstrating that the sequencing of interventions matter. This corresponds with findings in \citet{Enfors13} where it was observed that conservation tillage increases productivity significantly more during good rainfall seasons than in dry periods. Another conclusion coming from the models is that because of severe nutrient limitations existing in the case-study catchment, like much of sub-Saharan Africa, interventions focusing on water technology will only be effective with simultaneous nutrient inputs. Thus, modeling results may highlight potential benefits or shortcomings even before an intervention or empirical experiment takes place and help in their design.


\section{Discussion}

Alleviating poverty in rural agricultural settings is particularly challenging because of the interdependence between economic well-being, agricultural practices and the state of the biophysical environment. Interventions that only address single aspects of one or the other and neglect the other dimensions are likely to lead to unintended or ineffective outcomes. We show how poverty and soil dynamics are deeply interlinked and jointly determine the ability to meet food security goals in rural areas. An intervention targeting economic well-being through improved agricultural productivity using artificial fertilizers will fail in an environment where soil quality is compromised. At the same time interventions to improve soil quality, e.g. through conservation tillage will be unsuccessful if initial soil quality and economic well-being are too low. The complex and dynamic nature of interactions means that a blanket solution for persistent poverty does not exist and a sequence of interventions, rather than only one intervention, may be necessary for escape from the trap (Table 2 in SI). Models such as the ones presented here can be useful tools to test implications of dynamic interactions between the different dimensions and to identify which sequences may be appropriate in different contexts.

Our work advances understanding of the complex dynamics of rural poverty by combining the neoclassical economic theory of growth, ecological theories of nutrient cycling and empirical knowledge of interventions and development strategies. In situations with persistent poverty simply improving agricultural practices is not enough. Instead, a careful assessment is needed of the current state of the social-ecological system, including the socio-economic conditions of households, the biophysical conditions of the agroecosystems such as soil quality, nutrient and water availability and existing agricultural practices. Based on an understanding of a given context, combinations of interventions can be devised. These will most likely have to include methods to improve economic and biophysical conditions as well as initiating  changes in farmers’ habits and agricultural practices. 

Our analysis gives three main insights for development practice (Table 2). First, agrochemical inputs can sometimes reinforce poverty by degrading soil quality. Because of this, monitoring soil quality and moderate use of agrochemicals are potentially good practices. Second, prioritising diversification of energy sources can establish the conditions for effective application of other strategies. This is however possible only if people change their habits of using manure as a fuel source. Third, sequencing of interventions matters for conservation tillage to be effective because it preserves existing but does not contribute additional nutrients and water. In cases where there is not enough nutrients or water, conservation tillage should be combined with nutrient or water inputs and eventually followed by asset inputs.

The theoretical models presented here serve as thinking tools to unravel the complex dynamics and context-dependence of poverty traps in rural areas. We have built them on a synthesis of insights from empirical research. Future work should directly test these models and implications with data-based empirical models. Furthermore, our models focus on the importance of biophysical dynamics for escaping poverty traps at the households scale. Since nitrogen is often a limiting factor for crop growth, studying its dynamics is an important research question. Its concentration in the soil can be increased by intercropping with nitrogen fixating plants. Our models can easily be extended to represent nitrogen dynamics, but it was beyond the scope of this paper and we leave it for the future research.

Poverty traps dynamics are, however, influenced by many factors at and across scales \citep{Haider}. Future research may thus include cross-scale effects caused by e.g. population structure, migration, or the relationship between urbanization and poverty \citep{Chen,DeBrauw,Hunter}. Another important aspect that we only touch upon is the need to consider human behavior \citep{Beckage} and culture \citep{Lade}.  

In summary, nutrients, water, soil quality and household assets are critical factors for agricultural productivity, and their interactions can lead to reinforcing or breaking poverty traps. Dynamical systems modelling, which we used here, enables the testing of assumptions across various contexts to examine the implications of different agricultural interventions for poverty alleviation. As our models demonstrate, effective poverty alleviation is often best achieved by a planned sequence of interventions, rather than just one strategy. 


{\bf Code availability:} The Mathematica code used to generate the state space plots with attractors and basins of attractons in this article is available upon request from the corresponding author. The algorithms for plotting two- and three-dimensional basins of attraction were originally developed by \cite{Lade}. 

{\bf Acknowledgements:} We are grateful to our colleagues Million Belay and Linus Dagerskog for providing comments and empirical background for the paper.  

{\bf Funding:} The research leading to these results received funding from the Sida-funded GRAID program at the Stockholm Resilience
Centre, the European Research Council (ERC) under the European Union’s Horizon 2020 research and innovation programme (grant agreement No 682472 — MUSES), and the Swedish Research Council Formas (project grant 2014-589).

\bibliographystyle{plainnat}
\bibliography{ref}

\end{document}


\maketitle

{\bf Assets dynamics.}
In the neoclassic economic theory of growth, income is generated using production function $Y=F(\mathbf{K},L)$, where $\mathbf{K}=(K_a,K_p,K_w,K_q)$ is the vector of assets, phosphorus, water and soil quality needed for production and $L$ is the labor. Income is used for consumption and savings, $Y=C+S$, and savings are invested in production to provide generating future income. If $s$ is the savings rate, than investment can be expressed as $sY$ and dynamics of household assets $K_a$ on the aggregate level is given by
\begin{align}\label{1}
\frac{dK_a}{dt}=sY-\delta_aK_a,
\end{align}
where $\delta_a$ is the assets depreciation rate.

To study poverty, it might be more useful to consider per capita assets $k_a$ instead of their aggregate levels. We define per capita variables as $k_a=K_a/L$, $y=Y/L$ and $f(\mathbf{k})=F(\mathbf{K}/L,1)$ and per capita population (labor) growth rate $r=\frac{dL}{L\,dt}$. Since we have that
\begin{align}\label{2}
y=\frac{Y}{L}=F(\mathbf{K}/L,1)=f(\mathbf{k}).
\end{align}
The rate of change of per capita assets is
\begin{align}\label{3}
\frac{dk_a}{dt}=\frac{1}{L}\frac{dK_a}{dt}-\frac{K_a}{L}\frac{dL}{L\,dt}=\frac{1}{L}\frac{dK_a}{dt}-rk_a.
\end{align}
Combining equations  (\ref{1}), (\ref{2})  and (\ref{3}) it follows that
\begin{align}
\frac{dk_a}{dt}=s(k_a)f(\mathbf{k})-(\delta_a+r)k_a.
\end{align}

Let $$F(\mathbf{K},L)=AK_a^{\alpha_a}K_p^{\alpha_p}K_w^{\alpha_w}K_q^{\alpha_q}L^{1-(\alpha_a+\alpha_p+\alpha_w+\alpha_q)}, \quad A>0, \quad \alpha_a+\alpha_p+\alpha_w+\alpha_q \le 1,$$
be a Cobb-Douglas production function. Using the above variable change, we get 
$$f(\mathbf{k})=f(k_a,k_p,k_w,k_q)=Ak_a^{\alpha_a}k_p^{\alpha_p}k_w^{\alpha_w}k_q^{\alpha_q}, \quad A>0, \quad \alpha_a+\alpha_p+\alpha_w+\alpha_q \le 1,$$
which leads us to equations (1)--(4) in the main text.

\newpage
\begin{figure}[ht]
\begin{minipage}[b]{0.3\textwidth} 
\begin{subfigure}[b]{\linewidth}
\includegraphics[width=\linewidth]{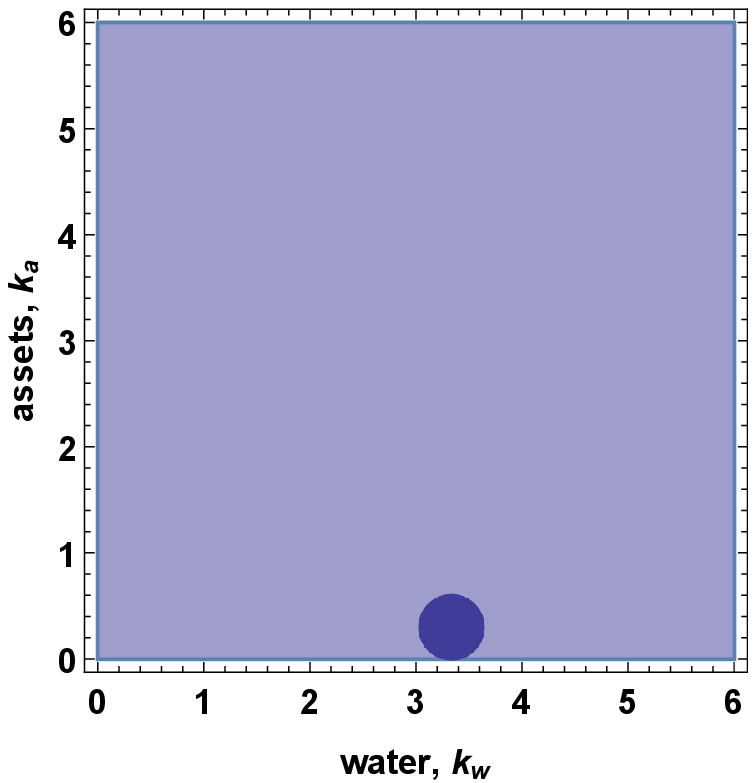}
\caption{}     
\end{subfigure}
\end{minipage}
\hspace*{0.5cm}
\begin{minipage}[b]{0.3\textwidth}
\begin{subfigure}[b]{\linewidth}
\includegraphics[width=\linewidth]{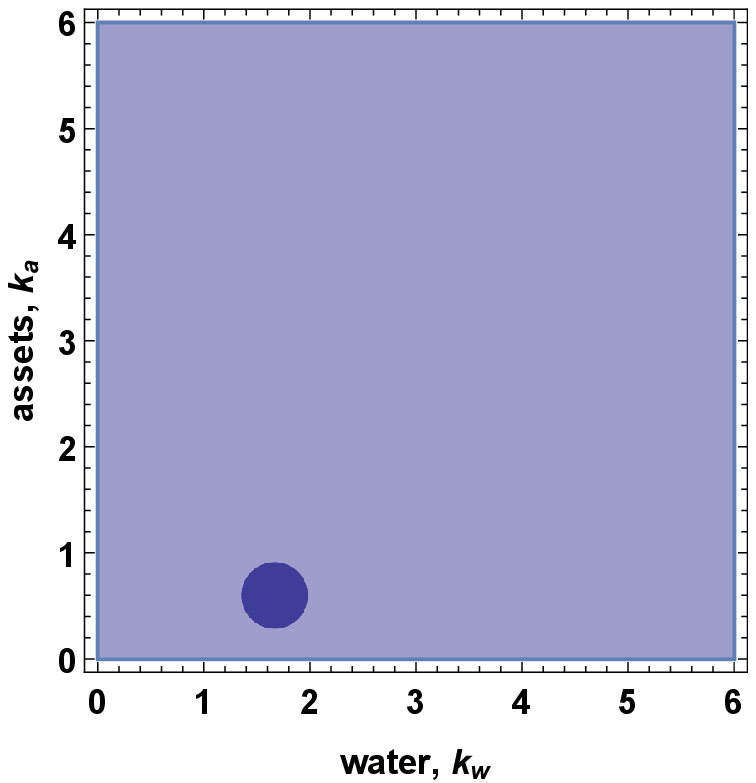}
\caption{}
\end{subfigure}
\end{minipage}
\caption{Conservation tillage. (A) Low phosphorus content of the soil and good water conditions, $A=3$. (B) High phosphorus level and poor water conditions, $r_w=0.5$.}
\label{Figure4}
\end{figure}


\maketitle

\begin{table}[h]\footnotesize
\begin{tabularx}{\textwidth}{m{2cm}|m{4cm}|m{3cm}|m{3cm}|m{3cm}}
\toprule
\bf{Model}  & \bf{Results} & \multicolumn{3}{l}{\bf{Insigths}} \\\midrule
The baseline model & Poverty regardless of water and soil conditions driven by phosphorus depletion (Figure 2) &  \multicolumn{3}{l}{\parbox{9cm}{Short term asset inputs are not useful because they do not change systems dynamics - a transformation of the system which would allow existence of several attractors is needed.}}
\\\midrule
Input of agrochemicals & Escape from poverty is possible under good water conditions and weak negative effect of agrochemicals (Figure 3B)
\newline
\newline
The harmful effects of agrochemicals on productivity via degradation of soil quality can outweigh improvements in productivity via improved soil nutrient levels, leading to a poverty trap  (Figure 3C) \vspace{0.2cm}
& \multicolumn{2}{c|}{\parbox{6cm}{ Agrochemical use may reinforce poverty through environmental degradation.
\newline
\newline
Under good water conditions and mild negative effects of agrochemicals, low asset level leads to poverty, but external asset input offers escape. }} & \multirow{3}{*}{\parbox{3cm}{Sequencing (a combination of several different strategies applied in a specific temporal order) is often necessary for poverty alleviation. 
\newline
\newline
The order of strategies is context dependent i.e. determined by biophysical properties, adopted practices and the initial state of a household.
\newline
\newline
Providing good water conditions is a necessary prerequisite for success of other interventions.
}} 
 \\\cline{1-4}
Diversification of household energy sources & \vspace{0.2cm}Escape from poverty is possible provided good water conditions and sufficient amount of nutrient rich manure (Figure 3E)  
\newline
\newline
Poverty is the only outcome if there is not enough manure or manure nutrient levels are low (Figure 3F) 
& \vspace{0.2cm} Escape from poverty is possible only after habit changes regarding household energy sources use.
\newline
\newline
In degraded environments, external phosphorus inputs prior to manure use might be needed.\vspace{0.2cm}
 & \multirow{2}{*}{\parbox{3cm}{Techniques to conserve nutrients are not useful in degraded environments because the total amount of nutrients present in the system sets the limit for crop production. 
\newline
\newline
Conservation techniques are useful when there is sufficient amount of nutrients in the system. 
}} &
 \\\cline{1-3}
Conservation tillage  &\vspace{0.2cm} Without external nutrient inputs, dynamics correspond to the baseline model because of the dominant negative phosphorus balance (Figure 2) 
\newline
\newline
With external nutrient inputs, escape from poverty is possible under good water conditions, but initial conditions play important role for the outcome.(Figure 3A-C)
\newline
\newline
If conservation tillage preserves phosphorus (water), non-poor equilibrium exists only for high phosphorus (water) levels. (Figure 4)
 & & &
 \\\bottomrule
\end{tabularx}
\smallskip
\caption{Main results and insights }\label{Table2}
\end{table}